# Lazy Evaluation: A Comparative Analysis of SAS MACROs and R Functions

Chen Ling[1] and Yachen Wang[1]

[1]AbbVie Inc., Lake County, Illinois, USA

Corresponding author: Chen Ling (ling0152@umn.edu)

## Abstract

Lazy evaluation is a powerful technique that can optimize code execution by deferring evaluations until their results are required, thus enhancing efficiency. In most modern programming languages, like R, lazy evaluation is commonly applied to function arguments. However, the application of lazy evaluation in SAS has not been extensively explored. This paper focuses on the mechanisms of lazy evaluation in SAS MACROs and R functions, offering a comparative analysis of the underlying principles that drive these processes.

R's lazy evaluation is driven by a data structure called Promise, which postpones evaluation and does not occupy memory until the value is needed, utilizing a call-by-need strategy. SAS, on the other hand, achieves lazy evaluation through its symbol tables, employing memory to store parameters, and operates on a call-by-name basis. These discrepancies in lazy evaluation strategies can notably impact the results of R functions and SAS MACROs. By examining these distinct approaches, the paper illuminates the impact of lazy evaluation on programming efficiency, supported by illustrative examples. As the shift from SAS to R becomes increasingly prevalent in the pharmaceutical industry, understanding these techniques enables programmers to optimize their code for greater efficacy. This exploration serves as a guide to enhance programming capabilities and performance in both languages.

## 1. Introduction

In data analysis and statistical programming, achieving efficiency and reducing computational overhead are very important. One method to enhance efficiency is through lazy evaluation—a technique that delays the evaluation of expressions until their results are actually needed. As one of the modern programming languages, R employs lazy evaluation for function arguments, leveraging a data structure known as Promises and a call-by-need evaluation strategy to delay computation until necessary. This approach enhances efficiency, reducing unnecessary resource usage and allowing for more streamlined processes. While R's implementation of lazy evaluation is well-recognized, the application within SAS MACROs remains less explored.

SAS has been a traditional choice in the pharmaceutical industry's programming toolkit for decades, renowned for its robust data analysis capabilities and macro functionality. Although SAS employs symbol tables for variable storage, which inherently supports deferred computations, it does so using a call-by-name strategy—a fundamentally different approach compared to R's call-by-need methodology.

Understanding these distinct mechanisms is essential, especially as many pharmaceutical companies are under transition from SAS to R (Ling & Wang, 2025). This paper seeks to illuminate the differences in lazy evaluation between SAS MACROs and R functions, offering insights and illustrative examples to guide programmers in optimizing their code. By delving into these nuances, programmers can leverage the strengths of each language and improve their programming efficacy.

## 2. Where do SAS and R store their parameters

A fundamental aspect of understanding lazy evaluation in programming languages like SAS and R is examining where and how parameters are stored. This storage mechanism directly influences how and when expressions are evaluated, affecting the efficiency and behavior of code execution.

### 2.1 Environments in R

R uses a memory area called an environment to manage variables and their values. The environment contains two main types: the global environment and the local environment, corresponding to the global and local scopes. In below, you'll see the source code on the left and a demonstration of the code execution on the right (Ling & Wang, 2025).

- In the first step, R creates two objects in the global environment: it assigns the value 6 to y, and a function to h (depicted with a little pink symbol).
- In the second step, function h() is invoked with h(1), which initiates an execution environment for h(). Within this environment, the variable a is set to 2, and x defaults to 1.
- In the third step, the function computes the sum of x and a, returning this result to be assigned to z in the global environment. Once the function h() completes its execution and returns the value, the execution environment is discarded.

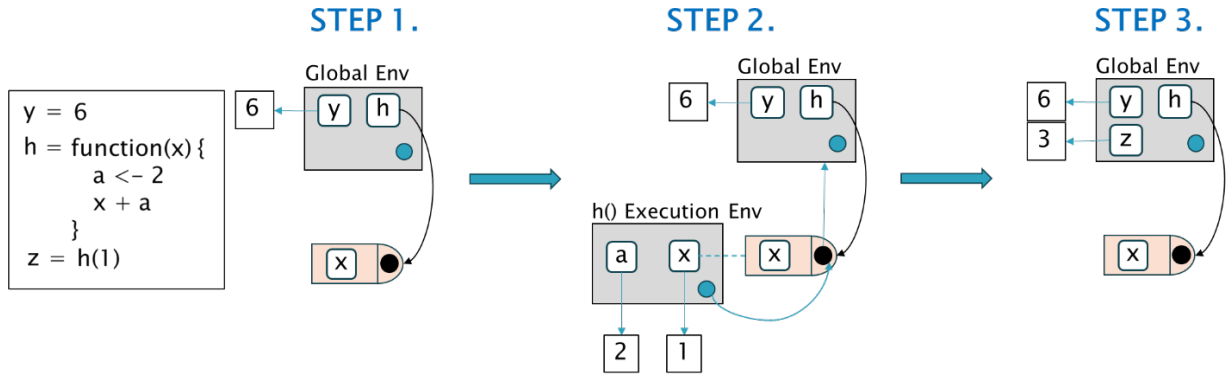

**Figure 1. Demonstration of environments in R**

## 2.2 Symbol tables in SAS

In SAS, variable scoping relies on a special memory area called the symbol table, which manages macro variables and is crucial for dynamic code substitution. Instead of the global and local environments, in R we have global and local symbol tables for global and local scopes. There is only one global symbol table created when the SAS session starts and is deleted when the SAS session ends. However, various local symbol tables are temporarily created during macro execution under specific conditions, then they will be deleted after the execution.

## 3. Lazy evaluation in R

Lazy evaluation means the arguments will be evaluated only if they are used. Implementing lazy evaluation in a programming language can enhance computation efficiency and reduce memory consumption, especially when dealing with very complicated computations, since the arguments will be evaluated only if needed. However, sometimes this would also make the codes more difficult for programmers to understand.

As shown in R program 1, the argument y depends on x, the argument z can even be determined by the variables (a and b) defined later in the function. The lazy evaluation can ensure that y is not evaluated at the time it is defined, using x = 5, instead the evaluation is performed later when y is called in the local environment by using x = 2.

```
###examples for lazy evaluation of arguments in R
lazy_eval<-function(x=5,y=x*10,z=a+b){
  x=2
  a=3
  b=4
  c(x,y,z)
}

lazy_eval()
[1]  2 20  7    ###Results in R
```

**R program 1 Lazy evaluation for arguments**

### 3.1 Promise

The lazy evaluation in R is based on a data structure called Promise (Wickham, 2019). A promise is a special type of object that represents an unevaluated function argument. It contains three key components:

- Expression: The actual R code or expression that is passed as the argument.
- Environment: The context in which the expression should be evaluated, capturing any variables or functions that are needed for evaluation.
- Value: Initially unset, this part stores the evaluated result of the expression once it has been computed.

When a function is called in R, each argument is wrapped in a promise rather than being immediately evaluated. This wrapping ensures that the expression is held off until its value is explicitly required in the function body. Once the value of a promise is needed, R evaluates the expression within its stored environment and the result is cached in the promise itself. If the promise is accessed again, R uses the cached result, avoiding redundant evaluation. The advantage of Promise data structure is that it does not occupy any memory spaces at the time of creation; it would only store the value after evaluation. This feature can enhance computation efficiency and reduce memory consumption, especially when dealing with very complicated computations.

However, promises cannot be manipulated with R code, it's more like the Schrodinger's cat in a quantum state, any attempt to inspect them with R code will force an immediate evaluation, making the promise disappear (Wickham, 2019). For better understanding of how promises powers lazy evaluation, refer to the demonstration in Figure 2.

Consider the example of y. When the function lazy_eval() is called, y is assigned with expression x*10, and a promise is initiated. The promise encapsulates both the expression x*10 and the environment function execution environment. Upon R's execution of the c(x,y,z), the promise y will be evaluated to get a value, the expression x*10 will be evaluated in the function execution environment where x has already been updated to 2, thereby resulting in y being computed as 20.

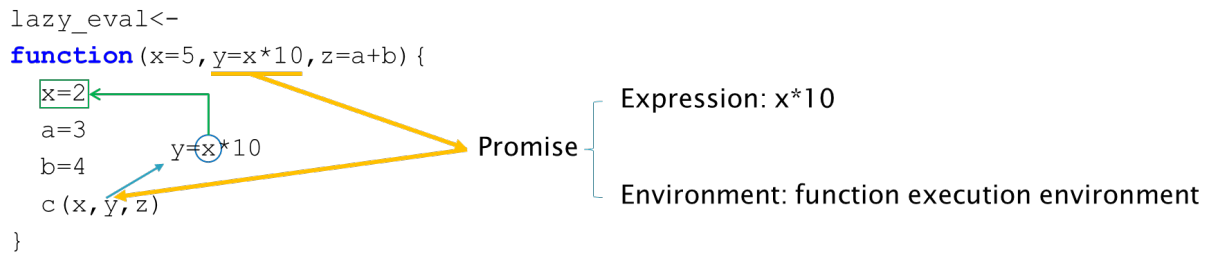

Figure 2. How promise data structure powers lazy evaluation in R

## 4. Lazy evaluation in SAS

SAS MACROs do not implement lazy evaluation to the default parameters in the same way that R does. However, if you try similar code in a SAS MACRO, you will get the same results as shown in SAS program 1, why is that?

```
/*examples for "lazy evaluation" of parameters in SAS*/
%macro lazy(x=5,y=&x*10,z=&a+&b);
%put _user_; /*This would display symbol tables in the log*/
%let x=2;
%let a=3;
%let b=4;
%put (&x %eval(&y) %eval(&z));
%mend;

%lazy()
(2 20 7)   /*Results in SAS log*/
```

SAS program 1. Lazy evaluation for parameters in SAS

The reason for getting the same results as lazy evaluation in R is because of the local symbol table in SAS (Ling & Wang, 2025). In SAS, as shown in Figure 3, the word scanner would read in the SAS codes, tokenize them and pass them to the compiler. When the word scanner detects a macro in the codes, it would automatically trigger a macro processor, which would create a local symbol table to store parameters as macro variables (Lyons, 2004). In our case, the local symbol table would store 5 as the value of x, '%eval(&x*10)' as the value of y and '%eval(&a+&b)' as the value of z.

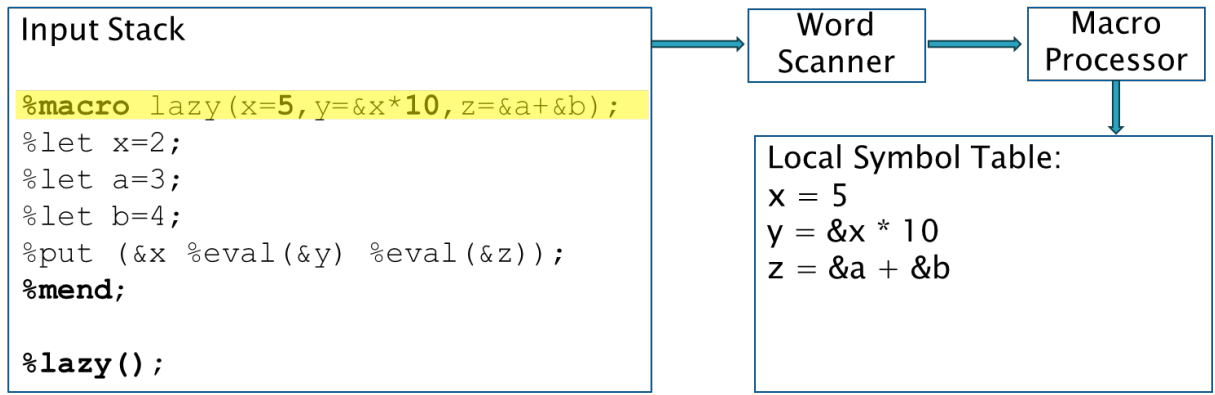

**Figure 3. Demonstration 1 of lazy evaluation in SAS**

Once the parameters in the macro statement are stored in the local symbol table, SAS proceeds to read the code inside of the macro, updating the value of x to 2 and adding a, b into the local symbol table. Then in %put (&x %eval(&y) %eval(&z))statement, SAS searches for and retrieves the value of x, y, z in the local symbol table, then writes the value into the log. For example, first %eval(&x * 10) is initially retrieved as the value of y, then the value of x will also be retrieved and substituted, resulting in %eval(2*10). In the following execution, value 20 will be put in the log as the resolution of y.

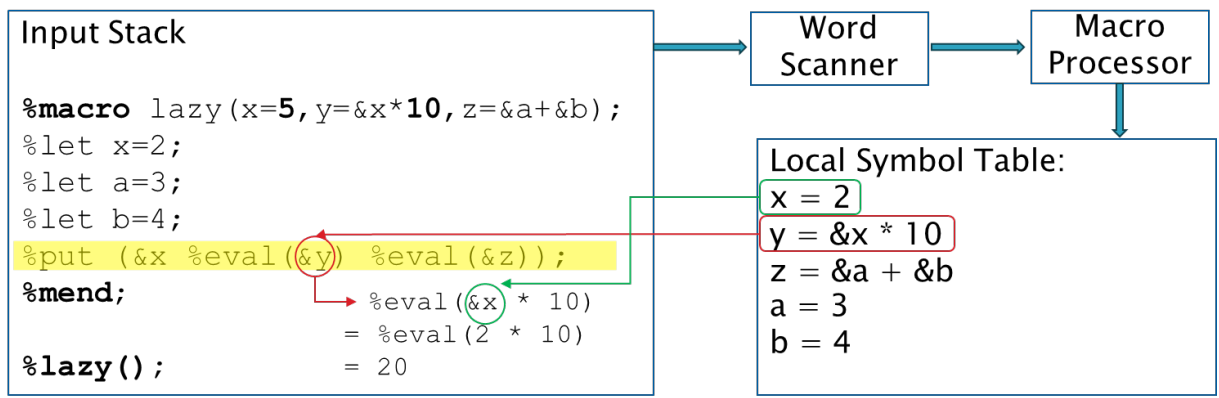

**Figure 4. Demonstration 2 of lazy evaluation in SAS**

R and SAS both use a place to store the expression of default arguments/ parameters, but Promises in R do not occupy memory until it's called, whereas SAS would use a memory space (Local Symbol Tables) to store the expressions.

## 5. Evalutaion Strategy

From the above examples, we learned that SAS and R are both using lazy evaluations to their default parameters, and would have the same results. However, due to the difference lazy evaluation mechanism, the evaluation strategies in SAS and R are not the same, which can occasionally result in different outcomes.

### 5.1 Call-by-Need in R

R employs a call-by-need strategy facilitated by its promise data structure. In this paradigm, expressions are not evaluated until their values are explicitly required, leading to a single evaluation cached for subsequent use. In R program 2, the value of y is printed twice with the same value. When it is first called within the print() function, the promise is evaluated using x = 2, returning 20 as the value to be printed. This value of the promise is then stored as the value of y. Thus, when the value of x in the environment is changed later, the value of y remains unchanged, as the promise is not re-evaluated.

```
lazy1<-function(x=5,y=x*10){
```

```
  x=2
  print(y)
  x=10
  print(y)
}
lazy1()

[1] 20
[1] 20
```

**R program 2. Example of call-by-need strategy in R**

This strategy enhances resource optimization by eliminating redundant computations, which is advantageous in data-intensive processing. It also allows R functions to conditionally execute code, ensuring that only necessary computations are performed. However, it necessitates a thoughtful design around promises to manage potential side effects and delayed error detection, as expressions are evaluated only when accessed.

### 5.2 Call-by-Name in SAS

In SAS, the call-by-name strategy is fundamental to its approach to lazy evaluation. With call-by-name, expressions are deferred until the moment they are required in the execution flow. This means that macro variables are evaluated in their textual form and substituted dynamically during runtime. In SAS program 2, similar to the R program 2, the value of y is put into the log twice. While the first output y is 20, matching the result in R, the second output changes to 100. This is because, with the call-by-name strategy, results from the first evaluation are not cached as they are in R. When SAS later executes x = 10, the value of x in the local symbol table is updated to 10. Consequently, the second %put statement re-evaluates y, resulting in y = 100 using x = 10.

```
%macro lazy1(x=5,y=&x*10);
%let x=2;
%put %eval(&y);
%let x=10;
%put %eval(&y);

%mend;

%lazy1()

20
100
```

**SAS program 2. Example of call-by-name strategy in SAS**

Although this approach provides flexibility in manipulating variable names and constructing adaptable code structures, it can lead to repetitive evaluations if variables are invoked multiple times. Consequently, programmers may need to carefully design their macros to balance flexibility and efficiency, ensuring that deferred evaluations do not introduce unnecessary overhead.

## 6. Conclusion

In this paper we compared the lazy evaluation mechanisms of SAS and R, highlighting key differences that underscore their respective strength. R employs call-by-need strategy, powered by the promise data structure, which defers computations and does not occupy memory until evaluation. This approach significantly enhances computational efficiency, reducing memory consumption and streamlining processes. In contrast, SAS employs a call-by-name strategy using symbol tables to enable dynamic and flexible macro functionalities, which are well-suited for complex data manipulation tasks. While this approach offers adaptability, it requires careful management to avoid redundant evaluations that could impact performance. As the pharmaceutical industry shifts from SAS to R, understanding these strategies equips programmers to leverage SAS's adaptability alongside R's efficiency. By embracing the unique strengths of both languages, developers can improve their coding practices, ensuring optimal performance and resource utilization across diverse analytical tasks.